\documentclass{optica-article}
\journal{opticajournal} 

\articletype{Research Article}

\usepackage{lineno}
\usepackage{siunitx}
\usepackage{color,soul}
\usepackage{lineno}
\usepackage{subfig}
\usepackage{amsmath} 
\usepackage{mathtools}
\usepackage{float}

\begin{document}

\pagenumbering{arabic} 

\title{Simultaneous measurement of multimode squeezing through multimode phase-sensitive amplification}


\author{Ismail Barakat, \authormark{1,$\dagger$,*} Mahmoud Kalash, \authormark{1,2,$\dagger$} Dennis Scharwald, \authormark{3} Polina Sharapova, \authormark{3} Norbert Lindlein, \authormark{1} Maria Chekhova \authormark{1,2}}

\address{\authormark{1}Friedrich-Alexander-Universität Erlangen-Nürnberg,Staudtstr. 7/B2, 91058 Erlangen, Germany\\
\authormark{2}Max-Planck Institute for the Science of Light, Staudtstr. 2, Erlangen D-91058, Germany\\
\authormark{3}Paderborn University, Department of Physics, Warburger Straße 100, D-33098 Paderborn, Germany}

\email{\authormark{*}Corresponding author: ismail.barakat@fau.de} 
\email{\authormark{$\dagger$} These authors contributed equally}

\begin{abstract*} 

Multimode squeezed light is an increasingly popular tool in photonic quantum technologies, including sensing, imaging, and computation. Meanwhile, the existing methods of its characterization are technically complicated, which reduces the level of squeezing, and mostly deal with a single mode at a time. Here, for the first time to the best of our knowledge, we employ optical parametric amplification to characterize multiple squeezing eigenmodes simultaneously. We retrieve the shapes and squeezing degrees of all modes at once through direct detection followed by modal decomposition. This method is tolerant to inefficient detection and does not require a local oscillator. For a spectrally and spatially multimode squeezed vacuum, we characterize eight strongest spatial modes, obtaining squeezing and anti-squeezing values of up to $-5.2 \pm 0.2$ dB and $8.6 \pm 0.3$ dB, respectively, despite the 50\% detection loss. 
This work, being the first exploration of OPA's multimode capability for squeezing detection, paves the way for the real-time detection of multimode squeezing.
\end{abstract*}

\section{Introduction}

Squeezed light plays a key role in photonic quantum technologies~\cite{Andersen2016Apr}. In metrology, it provides the measurement of the phase beyond the classical sensitivity limit~\cite{Giovannetti2004Nov}; its use in gravitational-wave detectors was crucial for the first observation of gravitational waves~\cite{Barsotti2018Dec,LIGOScientificCollaborationandVirgoCollaboration2017Oct}. Squeezed light enhances signal-to-noise ratio in Raman spectroscopy \cite{deAndrade2020May}, as well as in imaging \cite{Brida2010Apr} and microscopy~\cite{Samantaray2017Jul,Casacio2021Jun}. Noteworthy, imaging and spectroscopy applications require squeezed light that is multimode in space/angle and time/frequency, respectively.  Multiple squeezed modes are also needed in continuous-variable quantum information processing and quantum communication, where each mode serves as an information carrier~\cite{Roslund2014Feb} and a large set of modes can be used for measurement-based quantum computation~\cite{Chen2014Mar,Larsen2019Oct,Cai2017Jun}. 

Therefore, an important task is to characterize squeezing in multiple spatial and temporal modes. The standard way to measure squeezing is through homodyne detection, where the squeezed light interferes with a strong coherent beam, called the local oscillator, which matches its frequency and angular spectrum. But in the case of multimode squeezed light, each mode requires an individually shaped local oscillator. In practice, this means that different modes are characterized one by one, with the local oscillator shaped differently each time~\cite{Roslund2014Feb,LaVolpe2020Apr}. It is also possible to characterize a certain mode after filtering it from the rest; but because squeezing is highly susceptible to  loss, the filtering has to be projective. Although a Gaussian spatial mode can be indeed filtered out losslessly using a single-mode fiber~\cite{Perez2015Nov}, filtering of higher-order spatial modes requires more complicated methods to minimize losses~\cite{Berkhout2010Oct,Gu2018Mar,Zhou2017Dec}. It is even more difficult to perform projective filtering of a frequency mode: the only existing method is through nonlinear frequency conversion~\cite{Ansari2018May}. However, none of these methods allows the characterization of all modes simultaneously.

Following a proposal by  Beck~\cite{Beck2000Jun}, several experimental works ~\cite{Janousek2009Jul,Armstrong2012Aug} used array detectors to simultaneously characterize multiple spatial modes. This method requires no spatial shaping of the local oscillator, but addresses individual modes through post-processing with mode functions applied to the collected data. However, this approach has limitations: it is not suitable for complex modes and detects the same quadrature for all modes at once. The latter is significant when detecting cluster states or entanglement, which generally requires different quadratures of different modes to be addressed simultaneously~\cite{Cai2017Jun}. Another approach also involves array detectors to analyze pixel-wise correlations after homodyne detection ~\cite{Cai2021May}. While this technique also enabled the simultaneous reconstruction of multimode squeezing, both methods share common limitations of homodyne detection: they are sensitive to detection losses, especially significant with array detectors, they are constrained by the local oscillator spectral bandwidth, and restricted by the electronic bandwidth of the homodyne detector.

Recently, it was shown that squeezing can be measured via direct detection using optical parametric amplification~\cite{Shaked2018Feb,Frascella2019Sep,Takanashi2020Nov,Nehra2022Sep,Kalash2023Sep}. 
Indeed, the mean number of photons at the output of a phase-sensitive optical parametric amplifier (OPA) amplifying a generalized quadrature $\hat{Q}_\phi$ and de-amplifying the conjugated quadrature $\hat{P}_\phi$, defined at the optical phase $\phi$~\cite{Leonhardt1995Jan}, is 
\begin{equation}
    \langle \hat{N}_\phi \rangle =  e^{2G}\langle\hat{Q}_\phi^2\rangle + e^{-2G}\langle\hat{P}_\phi^2\rangle - \frac{1}{2},
    \label{eq:N}
\end{equation} 
where $G$ is the amplification gain.
If the amplification is sufficiently strong and the input state has $\langle \hat{Q}_\phi\rangle=0$, the output mean photon number becomes

\begin{equation}
\langle \hat{N}_\phi \rangle =  e^{2G}\Delta Q_\phi^2,
\label{eq:var}
\end{equation}
where $\Delta Q_\phi^2$ is the input quadrature variance. In other words, high-gain phase-sensitive amplification maps the variance of an input quadrature to the output mean photon number, or mean intensity (See SI, Sec.1, for more details). The quadrature that gets amplified can be selected by scanning the phase of the OPA pump. To find out the degree of squeezing of the quadrature, its variance is compared to the vacuum noise level, the latter can be obtained by amplifying the vacuum, i.e., blocking the amplifier input. 

This method of squeezing measurement offers several significant advantages. As shown previously, it can tolerate detection losses, supports broadband detection since OPAs are inherently broadband devices, and employs direct detection, which is often technically simpler than homodyne detection taking into account that the squeezer and the amplifier can be built using the same tools. Moreover, it is not restricted by the electronic bandwidth of homodyne detectors. On top of that, a traveling-wave OPA is a multimode device, which enables its use for noiseless amplification of images~\cite{Choi1999,Mosset2005}. Importantly, the eigenmodes of an OPA are, in general, complex~\cite{Beltran2017Mar,Frascella2019Oct,Sevilla-Gutierrez2024Feb}. Furthermore, an OPA can allow for the amplification of different quadratures across various input modes simultaneously via proper engineering of its modal structure. These features render OPA a powerful tool for multimode squeezing detection. Although achieving sufficient gain for OPA-based squeezing detection may be challenging for some experiments, the effort is justified by the significant benefits it provides.
\begin{figure}[h]
 \centering
 \includegraphics[width=0.9\linewidth]{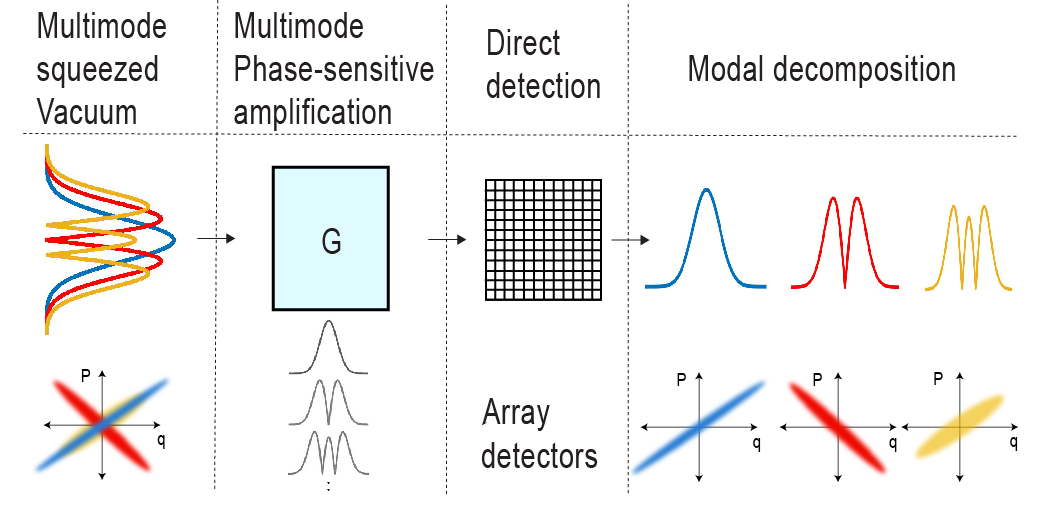}
\caption{Generalized and simplified scheme of measuring multimode squeezing through phase-sensitive optical parametric amplification.}
\label{Schematic}
\end{figure}

Here, for the first time to the best of our knowledge, we employ OPA to simultaneously characterize several modes of multimode squeezed vacuum. We combine the multimode nature of OPA with the method of reconstructing the shapes and weights of its eigenmodes~\cite{Finger2017May,Beltran2017Mar} to retrieve the squeezing of different modes simultaneously. Compared to previous works dealing with pulsed multimode squeezing \cite{Roslund2014Feb,LaVolpe2020Apr,Kouadou2023Aug,Roman-Rodriguez2023Jun}, our work shows superior performance in both the results and the simplicity of measurement. Thus, the method promises to boost high-dimensional quantum information processing based on squeezed light.

The technique is based on the fact that the Schmidt modes (also known as squeezing modes ~\cite{Wasilewski2006Jun}, broadband modes ~\cite{Eckstein2008Aug}, or supermodes ~\cite{Patera2010Jan}) of the signal and idler beams of an OPA coincide with the coherent modes of only one (signal or idler) beams taken separately~\cite{Averchenko2020Nov}. Furthermore, given that each mode is populated by a thermal state, the modal decomposition can be reconstructed by measuring the covariance of intensities at the OPA output. The measured weights define the mean photon numbers in these modes. According to Eq.~(\ref{eq:var}), if the OPA modes are seeded with squeezed vacuum, these weights also define the squeezing levels. It is worth mentioning that the modal content of the OPA is defined by two factors,  the OPA phase-matching function and the pump profile~\cite{Wasilewski2006Jun,Eckstein2008Aug,Patera2010Jan}. Thus, a set of arbitrary input modes can be matched by properly engineering one or both these factors.~\cite{Patera2012Sep}

Figure \ref{Schematic} illustrates the general scheme of our method in the case of spatial modes. A multimode squeezed vacuum (or any multimode state, where each mode $m$ exhibits zero mean quadratures, $\langle \hat{Q}_m\rangle=\langle \hat{P}_m\rangle=0$), is directed to a multimode OPA that aligns with the input mode structure. By changing the OPA pump phase, we scan the phase $\phi$ of the amplified quadrature simultaneously across all modes. 
Subsequently, an ensemble of far-field intensity distributions is measured at the output, from which the mean photon number of each spatial mode is extracted.

\section{Parametric amplification of multimode squeezed vacuum}
\begin{figure}[h]
\centering
\includegraphics[width=0.8\linewidth]{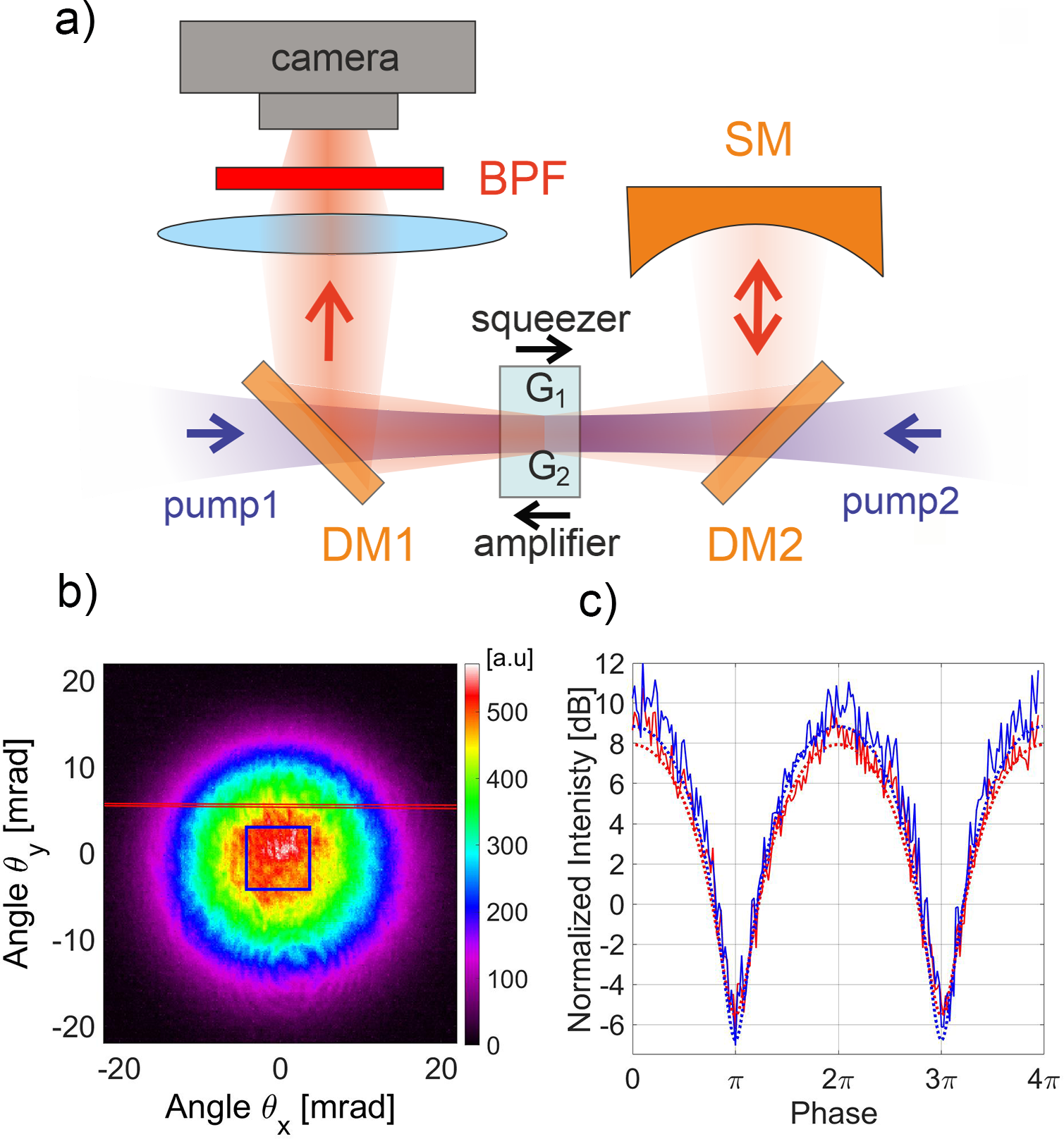}
\caption{(a) Simplified scheme of the setup. The same nonlinear crystal acts as a squeezer (left to right, pump 1) and an amplifier (right to left, pump 2). Multiple modes of the squeezed vacuum (SV) are imaged from the squeezer to the amplifier by means of spherical mirror SM. A camera registers far-field intensity distributions after a band-pass filter (BPF). (b) Intensity distribution registered by the camera. The blue square and the red stripe indicate the areas used for measuring squeezing in the collinearly emitted SV and for spatial modes reconstruction, respectively. (c) The intensities at the amplifier output integrated over the blue square (blue trace) and the red stripe (red trace) in panel (b), normalized to their values in the case of vacuum at the input, plotted against the pump 2 phase. The minima (maxima) indicate the degree of squeezing (anti-squeezing). Dotted lines show the results of theoretical calculations.}
\label{SqEffandAI}
\end{figure}

We generate multimode squeezed vacuum (SV) in a nonlinear beta-barium borate (BBO) crystal through type-I collinear frequency-degenerate parametric down-conversion (PDC), pumped by $18$ ps pulses at 354.67 nm (pump 1), see Fig.~\ref{SqEffandAI}a. The squeezing parameter for collinear emission at the degenerate wavelength 709.34 nm is $G_1 = 1.1\pm0.3$. The multimode SV is then reflected by dichroic mirror DM2 to a spherical mirror (SM), which images each spatial mode onto the same nonlinear crystal. Now, the crystal acts as a phase-sensitive
amplifier. It is pumped by pump 2, coherent and synchronized with pump 1. Pump 2 has the same beam waist $70\,\mu$m on the crystal but a higher energy per pulse, which leads to a larger parametric gain, $G_2 = 4.0\pm0.4$ for the collinear direction. Whether amplification or de-amplification occurs, is determined by the phase of pump 2, which is controlled with a piezoelectric actuator. For more details of the experimental setup, see the SI, Secs. 2,3.

After undergoing phase-sensitive amplification,  multimode SV is reflected by dichroic mirror DM1 and filtered by band-pass filter BPF with $10$ nm bandwidth centred at the degenerate wavelength. Finally, a gated and cooled sCMOS camera with \(55\%\) quantum efficiency, positioned at the Fourier plane of a lens, acquires 2D intensity distributions.

Figure~\ref{SqEffandAI}b displays the distribution averaged over $1500$ pulses. We align the spherical mirror so that while scanning the phase of the propagating pump beam in the second arm, the intensity distribution oscillates as a whole~\cite{Frascella2019Sep}. This procedure is aimed at balancing the phases of all spatial modes so that the condition for both amplification and de-amplification is met concurrently across all modes. To demonstrate squeezing for the collinearly emitted SV (Fig.~\ref{SqEffandAI}c), we integrate the signal over the central area of the camera (the blue square in Fig.~\ref{SqEffandAI}b). This integrated signal is then normalized to the signal measured when the radiation of the squeezer is blocked, allowing the OPA to amplify solely the vacuum. The measurement shows a visibility of $94\pm1$\(\%\) and the degrees of squeezing and anti-squeezing of {$-6.0\pm 0.6$} dB and $10\pm1$ dB, respectively. These numbers match the theoretical expectations, with an account for the losses of $15\%$, with $2\%$ measured optical losses and $13\%$ estimated losses due to modes mismatch between the squeezer and the amplifier  (see Sec. \ref{schmidt} and the SI, Secs. 3,4). 

Further, we acquire three separate sets of single-pulse frames: (1) with the phase locked at the `bright fringe', corresponding to the maximum of the interferogram in Fig.~\ref{SqEffandAI}c and to the amplification of the anti-squeezed quadrature; (2) with the phase locked at the `dark fringe', corresponding to the minimum of the interferogram and to the amplification of the squeezed quadrature; (3) for the amplified vacuum. Each dataset comprises $1500$ frames and is further used to extract the shapes and weights of the spatial eigenmodes. Together with the integral mean intensity, the weight of each mode determines its mean photon number after the OPA, and so, according to Eq.~\ref{eq:var}, the variance of the amplified quadrature. 

\section{Schmidt modes reconstruction}\label{schmidt}
The Schmidt modes of a SV are those minimizing the number of correlations between its signal and idler subsystems~\cite{Braunstein2005May}. 
In particular, the Schmidt modes of type-I collinear degenerate PDC very closely resemble the Hermite-Gauss modes of paraxial beams~\cite{Straupe2011Jun, Averchenko2020Nov}. Although these modes are two-dimensional  and can be represented as functions of both $\theta_x$ and $\theta_y$ angles in the far field, for type-I collinear PDC they can be considered, to a good approximation, as factorable: $u_{mn}(\theta_x,\theta_y)=u_m(\theta_x)u_n(\theta_y)$~\cite{Straupe2011Jun, Averchenko2020Nov}. For simplicity, we focus solely on one-dimensional modes $u_m(\theta_x)$. To find them, we rely on their identity with the coherent modes of one subsystem, signal or idler~\cite{Frascella2019Oct,Averchenko2020Nov}. We filter out one subsystem by considering only the upper part of the far-field intensity distribution depicted in Fig.~\ref{SqEffandAI}b (for more detail, see SI, Secs. 5,6). Alternatively,  signal-idler cross-correlations can be eliminated by non-degenerate frequency filtering~\cite{Averchenko2020Nov} 

The one-dimensional modes are extracted by processing single-pulse one-dimensional intensity distributions $I(\theta_x)$. To obtain them, we select a stripe covering the $\theta_y$ values from $\theta_{1}=6.5$ mrad to $\theta_{2}=7.5$ mrad (red lines in Fig.~\ref{SqEffandAI}b) and integrate over $\theta_y$:  $I(\theta_x)\equiv\int_{\theta_1}^{\theta_2}I(\theta_x,\theta_y)d\theta_y$.
We note that  because of the shift of the selected stripe from the collinear direction, both squeezing and antisqueezing is reduced compared to the one for the collinear direction. This is clear from   Fig.~\ref{SqEffandAI}c, where the red line shows the intensity integrated over the red stripe as a function of the amplification phase. Effectively, the gain is reduced to $G_1=1.05$.  

\begin{figure}[h]
\includegraphics[width=1\linewidth]{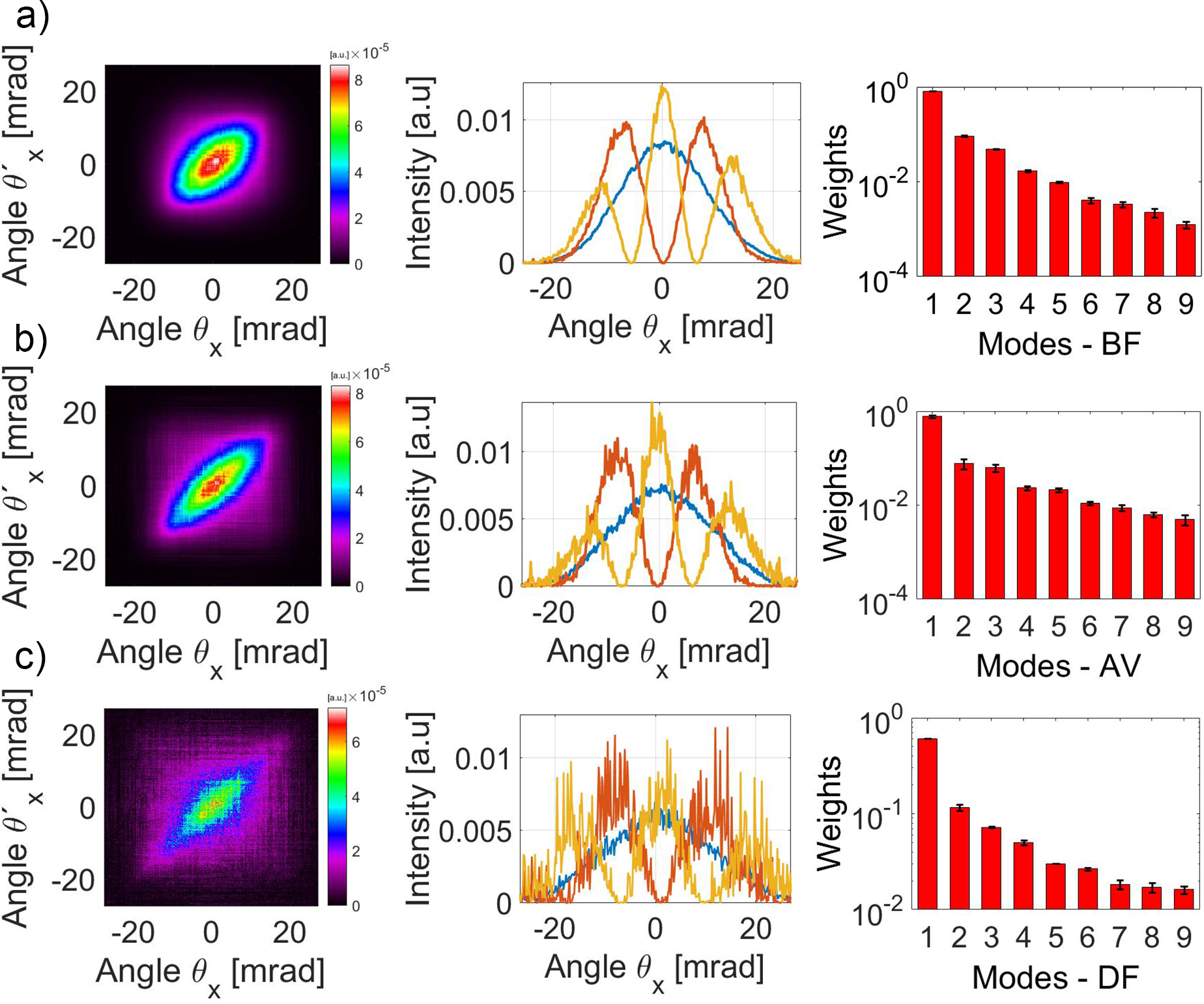}
\centering
\caption{From left to right: the normalized covariance $C(\theta_x,\theta'_x)$, the normalized intensity distributions of the three lowest-order modes, and the weights of the modes for the cases where OPA amplifies the anti-squeezed quadrature of the SV ('bright fringe', a), the vacuum (b), and the squeezed quadrature of the SV ('dark fringe', c).}
\label{3x3Figures}
\end{figure}

Further, from single-shot one-dimensional intensity distributions $I(\theta_x)$ we  calculate the intensity covariance,
\begin{equation}
    Cov({\theta_x},{\theta'_x}) = \langle I(\theta_x)I(\theta'_x)\rangle - \langle I(\theta_x)\rangle \langle I(\theta'_x)\rangle.
\end{equation}
The resulting normalized  
covariance distribution  $C(\theta_x,\theta'_x)=Cov(\theta_x,\theta'_x)/\int d\theta_x d\theta'_x Cov(\theta_x,\theta'_x)$ (Fig.~\ref{3x3Figures}, left panels) shows spatial auto-correlations in the signal beam ~\cite{Averchenko2020Nov}. Different panels show the cases where the anti-squeezed (a, `bright fringe'), squeezed (c, `dark fringe') quadratures  of the SV and the vacuum (b) are amplified. In all cases, the covariance distribution 
contains the modes of only the signal subsystem. We stress that the amplifier gain should be sufficient to provide an acceptable signal-to-noise ratio when amplifying the squeezed quadrature. Otherwise, the covariance distribution becomes noisy and so do the results.

As shown in Refs.~\cite{Beltran2017Mar,Frascella2019Oct}, the normalized covariance can be  written as (see SI, Sec.~5),
\begin{equation}
C(\theta_x,\theta'_x) = \left[ \sum_{m} \lambda_{m} u_{m}(\theta_x) u_{m}^*(\theta'_x) \right]^2, \label{eq_NC}
\end{equation}
where $\lambda_{m}$, $\sum_{m} \lambda_{m} = 1$, are the eigenvalues (weights) of the modes. These weights determine the populations $\langle \hat{N}_m \rangle$  of  different modes relative to the total mean photon number $\langle \hat{N}_{\mathrm{tot}} \rangle$~\cite{Perez2015Nov},
\begin{equation}
\langle \hat{N}_m \rangle =\lambda_{m}\langle \hat{N}_{\mathrm{tot}} \rangle.
\label{Nm}
\end{equation} 
Applying the singular-value decomposition to $\sqrt{C(\theta_x,\theta'_x)}$ yields the weights and shapes of the modes. The middle panels of Fig.~\ref{3x3Figures} show the intensity distributions of the three strongest modes, $|u_{m}(\theta_x)|^2$, while the right-hand panels a,b,c depict the mode weights $\lambda_m^{\mathrm{B}}$, $\lambda_m^{\mathrm{AV}}$, and $\lambda_m^{\mathrm{D}}$ for the bright fringe, amplified vacuum, and dark fringe, respectively. Further, we cut the  $\lambda_m$ distribution and apply the normalization to $12$ modes, as the reconstruction becomes increasingly noisy for higher-order modes.

The weight of the first mode is anomalously high because of the pump intensity fluctuations (see SI, Sec.6). Apart from that, the effective number of modes for the dark fringe is the highest and for the bright fringe, the lowest. Indeed, as the brightness of an OPA output increases, its effective number of populated modes decreases ~\cite{Sharapova2020Mar}. Nevertheless, even for the bright fringe, at least first eight modes are sufficiently populated to be characterized. 

In the general case, the Schmidt modes of the squeezer do not coincide with those of the whole two-crystal interferometer, which we reconstruct at the output. 
Even if the crystal lengths and the pump beam waists for the squeezer and amplifier are the same, their mode widths are different because they depend on the parametric gain and $G_1<G_2$. 
 Therefore, in the general case, each mode of the squeezer has a nonzero overlap with all modes of the OPA, as well as of the whole two-crystal interferometer, whose effective gain values are $G_1+G_2$ for the bright fringe and $G_2-G_1$ for the dark fringe. 
 This is why, to determine the degree of squeezing for a specific squeezer mode, one needs the weights of several (ideally, all) modes of the interferometer. It should be noted that the difference in sizes between the modes of the interferometer output for different cases (dark fringe, amplified vacuum, bright fringe) is not significant, as we can see in the middle panels of Fig.~\ref{3x3Figures}. This is because the corresponding effective gain values ($G_2-G_1=3,G_2=4,G_2+G_1=5$) are close. However, the mode sizes of the squeezer are significantly different from those of both the amplifier and the interferometer.

\begin{figure}[h!]
\centering
\includegraphics[width=1\linewidth]{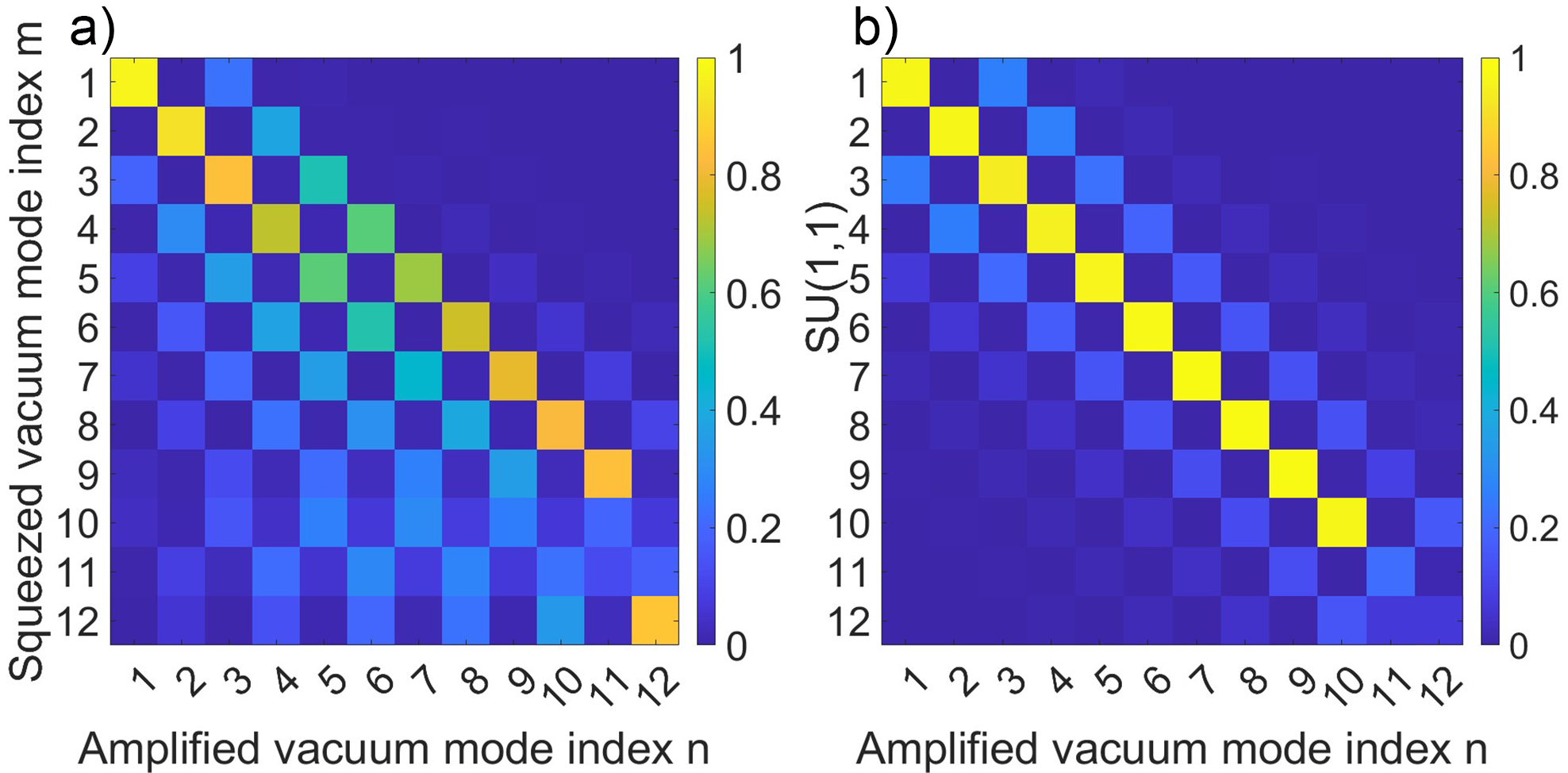}
\caption{The overlap coefficients: (a)  $g_{ln}$ between the modes of the squeezer and the input modes of the amplifier and (b)  $h_{ln}$  between the modes of the amplifier output and the whole interferometer in the bright fringe. }
\label{ModesMatching}
\end{figure}

Figure~\ref{ModesMatching}a shows the overlap matrix between the modes of the squeezer $v_l(\theta_x)$ and the input modes of the amplifier $w_n(\theta_x)$: $g_{ln} = \int d\theta_x \left[ v_l (\theta_x)\right]^*  w_n (\theta_x) $, normalized as $\sum_n |g_{ln}|^2 = \sum_l |g_{ln}|^2 =1$. Since we have sufficient theoretical knowledge about our source and OPA modal contents due to numerous previous studies, the overlap was calculated numerically with the use of the experimental parameters.  Generally, the mode content can be fully reconstructed using mode-diagnosis techniques, for instance~\cite{Straupe2011Jun}. The `checkerboard' structure arises from the zero overlap between odd and even modes, which are even and odd functions of $\theta_x$, respectively. This pattern holds for modes up to order 9 and gets violated for higher-order modes. 

Noteworthy, it is possible to make the overlap matrix $g_{ln}$ nearly diagonal by focusing the pump beams in the squeezer and the amplifier differently. In the amplifier, whose gain is higher than that of the squeezer, the modes are broader. However, their angular profiles can be reduced by increasing the pump beam waist (SI, Sec. 7). While this is sufficient to  match the mode sizes in the case where the squeezer and amplifier have different gain, in general, proper engineering of the pump is required in the case of arbitrary complex squeezed modes.

Meanwhile, the overlap between the amplifier output modes $u^{\mathrm{amp}}_n (\theta)$ and the output modes of the two-crystal interferometer $u_k (\theta)$ is given by the matrix $h_{nk} = \int d\theta  \ \left[u^{\mathrm{amp}}_n (\theta)\right]^*  u_k (\theta) $
, with the normalization condition $\sum_n |h_{ln}|^2 = \sum_l |h_{ln}|^2 =1$ (Fig.~\ref{ModesMatching}b). As expected, the  $h_{ln}$ matrix is nearly diagonal; therefore, further on we assume $h_{ln}=\delta_{ln}$, where $\delta_{ln}$ is the Kronecker symbol.   

\section{Multimode squeezing measurement}
The multimode amplifier transforms each output mode of the squeezer into a set of modes at the output of the interferometer. Denoting the photon annihilation and creation operators of the squeezer output modes as $\hat{A}_l, \hat{A}_l^\dagger$ and of the interferometer output modes as $\hat{A}^{\mathrm{int,out}}_n, \left[\hat{A}^{\mathrm{int,out}}_n\right]^\dagger$, with $l,n$ numbering the modes 
(see Appendix A), we write the inverse multimode Bogoliubov transformation (\ref{eq:connection_relation}) as
\begin{equation} \label{eq:connection_App}
 \hat{A}_l = \sum_{n} g_{ln} \left[\sqrt{\Lambda_n+1}  \hat{A}_n^{\mathrm{int,out}} \mp  \sqrt{\Lambda_n } \left[\hat{A}_n^{\mathrm{int,out}}\right]^{\dagger} \right],
 \end{equation}
 where $\Lambda_n=\sinh^2(r_n)$, with $r_n$ being the parametric gain for mode $n$ of the amplifier, is the photon number in mode $n$ of the amplified vacuum. The minus sign here corresponds to the bright fringe, and the plus sign, to the dark fringe. 
 
From this relation, we can express the quadrature operators $\hat{Q}_l$ and $\hat{P}_l$ of the squeezer mode $l$ in terms of the quadrature operators $\hat{Q}^{\mathrm{B},\mathrm{D}}_n,\hat{P}^{\mathrm{B},\mathrm{D}}_n$ of the interferometer output modes $n$ in the bright (B) and dark (D) fringes. We assume, without the loss of generality, that the position quadratures $\hat{Q}_l$ are anti-squeezed while the momentum quadratures $\hat{P}_l$ are squeezed. 

We align the spherical mirror so that the intensity distribution at the amplifier output (Fig.~\ref{SqEffandAI}b) oscillates as a whole, with the intensity at each point being minimal for the dark fringe and maximal for the bright fringe. As shown in Appendix B, this procedure balances the phases of the modes, ensuring that the $g_{ln}$ matrix is real and that the position operator of each squeezed vacuum mode is related only to the  position operators of the amplifier output modes, and similarly for the momentum operators:
\begin{equation}
\label{eq:simpB}
\begin{split}
 \hat{Q}_l = \sum_{n} g_{ln}(\sqrt{\Lambda_n+1}-/+ \sqrt{\Lambda_n}) \hat{Q}^{\mathrm{B/D}}_n,
 \\
 \hat{P}_l = \sum_{n} g_{ln}(\sqrt{\Lambda_n+1}+/-\sqrt{\Lambda_n}) \hat{P}^{\mathrm{B/D}}_n.
 \end{split}
 \end{equation} 

Taking into account the absence of correlations between the quadratures of different Schmidt modes, we obtain the variances of the anti-squeezed and squeezed quadratures in each mode:
 \begin{equation}
\label{eq:Var1}
\begin{split}
 \Delta Q^2_l = \sum_{n} g^2_{ln}(\sqrt{\Lambda_n+1}-\sqrt{\Lambda_n})^2 \langle({Q}^\mathrm{B}_n)^2\rangle,
  \\
  \Delta P^2_l = \sum_{n} g^2_{ln}(\sqrt{\Lambda_n+1}-\sqrt{\Lambda_n})^2 \langle({Q}^\mathrm{D}_n)^2\rangle.
  \end{split}
 \end{equation} 
Provided the amplification is strong, the mean photon number in each mode is equal to the mean square of the amplified quadrature for the same mode. Then (see Fig.~S1), in the bright fringe $\langle N_n^{\mathrm{B}}\rangle=\langle({Q}^\mathrm{B}_n)^2\rangle $, while in the dark fringe, $\langle N_n^{\mathrm{D}}\rangle=\langle({P}^\mathrm{D}_n)^2\rangle $.
 
Meanwhile, for each vacuum mode, a similar equation can be written: 
 \begin{equation}
  \Delta Q^2_{vac} = (\sqrt{\Lambda_n+1}-\sqrt{\Lambda_n})^2 \langle N_n^{ \mathrm{AV}}\rangle,
 \end{equation} 
where $\langle N_n^{ \mathrm{AV}}\rangle = \langle({Q}^\mathrm{AV}_n)^2\rangle $ is the mean number of photons in the $n$th mode of the amplified vacuum.

Finally, we get
  \begin{equation}
\label{eq:Var}
 \Delta Q^2_l = \Delta Q^2_{\mathrm{vac}}\sum_{n} g^2_{ln} \frac{\langle N_n^{\mathrm{B}}\rangle}{\langle N_n^{\mathrm{AV}}\rangle},\,\Delta P^2_l = \Delta Q^2_{\mathrm{vac}}\sum_{n} g^2_{ln}\frac{\langle N_n^{\mathrm{D}}\rangle}{\langle N_n^{\mathrm{AV}}\rangle}.
 \end{equation} 
 
Applying Eq. (\ref{Nm}) to all three cases (bright fringe, amplified vacuum, dark fringe), we obtain the degrees of anti-squeezing  $ AS_l$ and squeezing $S_l$ in each mode in terms of the calculated overlap coefficients $g_{ln}$, measured mean total number of photons $\langle N^\mathrm{B}_{\mathrm{tot}}\rangle, \langle N^{\mathrm{AV}}_{\mathrm{tot}}\rangle, \langle N^\mathrm{D}_{\mathrm{tot}}\rangle$ and mode weights $\lambda_{n}^\mathrm{B}, \lambda_{n}^{\mathrm{AV}}, \lambda_{n}^\mathrm{D}$ for the bright fringe, amplified vacuum, and dark fringe, respectively:
 \begin{equation}
 \label{eq:result}
  AS_l= 10 \log_{10} \sum_{n} g^2_{ln}\left[ \frac{\lambda_{n}^\mathrm{B}\langle N^\mathrm{B}_{tot}\rangle}{\lambda_{n}^{\mathrm{AV}}\langle N^{\mathrm{AV}}_{tot}\rangle} \right],\,
 S_l= 10 \log_{10} \sum_{n} g^2_{ln}\left[ \frac{\lambda_{n}^\mathrm{D}\langle N^\mathrm{D}_{tot}\rangle}{\lambda_{n}^{\mathrm{AV}}\langle N^{\mathrm{AV}}_{tot}\rangle} \right].
  \end{equation}

\begin{figure}[h!]
\includegraphics[width=1\linewidth]{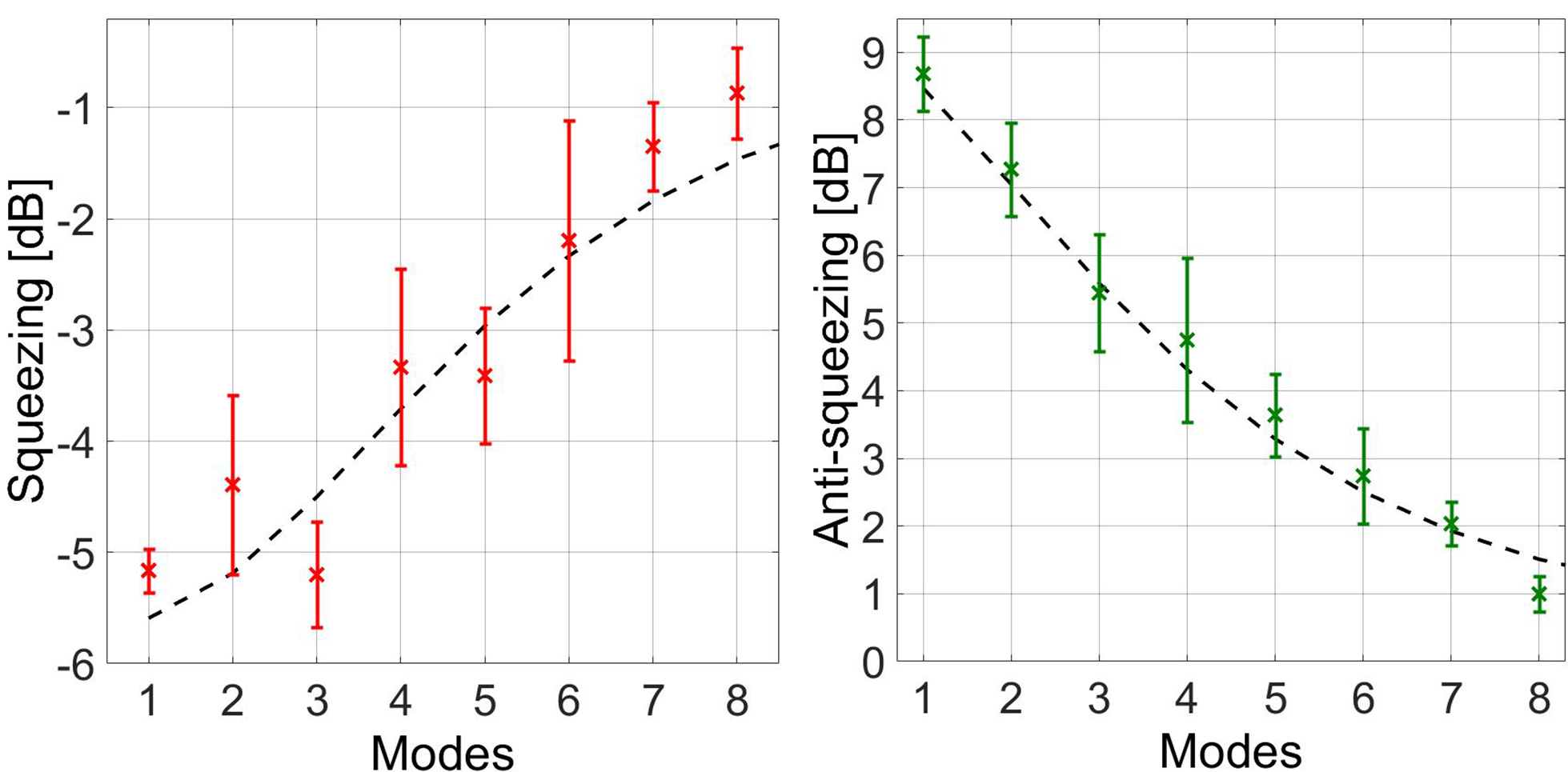}
\centering
\caption{The degrees of squeezing (left) and anti-squeezing (right), measured  (red and green points, respectively) and calculated (dashed black line).}
\label{Sq-modes}
\end{figure}

Figure~\ref{Sq-modes} shows the degrees of squeezing (left) and anti-squeezing (right) measured for eight strongest modes. Both the mean values and the uncertainties were obtained by analyzing four different sets of data. The highest values are measured for the first mode, with $-5.2\pm0.2$ dB squeezing and $8.6\pm0.3$ dB anti-squeezing. To the best of our knowledge, this is the strongest reported squeezing for multimode detection through optical parametric amplification. 
 These values agree well with the ones theoretically calculated for the output of the squeezer, assuming a parametric gain of $G=1.05$ and the overall losses amounting to $15\%$. The theoretical values of squeezing and antisqueezing for the same modes are represented, for convenience, by black dashed curves. The large uncertainty in the squeezing values for the highest-order modes is caused by the low level of the measured signal in the dark fringe and, as a consequence, by the inaccurate reconstruction of the modes shapes and weights. Characterization of higher-order modes is impossible, both for this reason and also because the parametric amplification for those modes is insufficiently strong to allow simplified relation~(\ref{eq:var}).

\section{Conclusion}

We have proposed and demonstrated a simple and efficient method to characterize multimode squeezed light. Our approach uses multimode optical parametric amplification followed by direct detection and mode weights reconstruction based on intensity correlations. In this method, all modes are measured simultaneously without the need for filtering any of them. In addition, the method is tolerant to detection inefficiencies and suitable for broadband and complex modes. As a proof of principle, we measured the degree of squeezing for eight strongest spatial modes of squeezed vacuum generated in a traveling-wave optical parametric amplifier (`squeezer'). We obtained the maximal values of squeezing and anti-squeezing of $-5.2\pm 0.2$ dB and $8.6\pm0.3$ dB, respectively. 

The results, showing stronger squeezing than in existing multimode experiments~\cite{LaVolpe2020Apr,Kouadou2023Aug,Roman-Rodriguez2023Jun}, are in a good agreement with theoretical predictions despite the imperfect overlap between the modes of the squeezer and the modes of the amplifier, caused by their different gain values. We account for the partial mode overlap by calculating it theoretically, based on our previous studies  of single-pass OPAs. This task is considerably simplified by balancing the phases between different modes, achieved by imaging the near field of the squeezer on the near field of the amplifier~\cite{Frascella2019Sep}. But the mode overlap between the squeezer and the amplifier can be made nearly perfect by properly engineering the OPA pump, which also allows one to retrieve  squeezing for  arbitrary complex modal content.

Here, we could not access modes of orders above 8 for several reasons: the invalidity of the high-gain condition of the amplifier, the low contributions of high-order modes into the overall intensity, leading to the impossibility to accurately balance their phases in experiment, and a high noise in their reconstruction, especially for the dark fringe. However, the approach can be extended to higher-order modes by designing an amplifier with more modes and a higher gain. 

This method can be 
easily extended to two-dimensional spatial modes as we showed in Ref.~\cite{Averchenko2020Nov}; in this case, signal-idler correlations should be eliminated by filtering a spectral band shifted from the degenerate wavelength. The method can be generalized to frequency modes and even mode combinations, to obtain and characterize cluster states for measurement-based quantum computation~\cite{Cai2017Jun}. 
It will thus support numerous high-dimensional quantum information experiments. Our findings also indicate that, if supplemented with mode sorting, the method can enable the simultaneous extraction of squeezing across multiple modes in real time, significantly enhancing the practicality of these squeezed light-based applications.

\section{Appendix A: Schmidt modes of the OPAs}
\label{App1}

Our theoretical description is based on solving systems of integro-differential equations for each crystal separately and for the entire system, forming an SU(1,1) interferometer. In this approach, the input-output relations (Bogoliubov transformations) for the plane-wave operators can be written as~\cite{Scharwald2023Nov}: 
\begin{equation}
    \hat{a}^{\mathrm{out}}(\theta) = \int d\theta' \eta(\theta,\theta') \hat{a}^{\mathrm{in}}(\theta')
                    + \int d\theta' \beta(\theta,\theta')
                    \left[\hat{a}^{\mathrm{in}}(\theta')\right]^{\dagger}.
        \label{eq:sol_ops_first_crystal}
             \end{equation}
Here, the transfer functions $\eta$ and $\beta$ can be decomposed using the joint Schmidt decomposition (Bloch-Messiah reduction) \cite{Christ2013May,Averchenko2020Nov}:  
            \begin{equation}
                \begin{split}
            \beta(\theta,\theta') = \sum_{n} \sqrt{\Lambda_{n}} u_{n}(\theta)w_{n}(\theta'), \\
                               \eta(\theta,\theta')  = \sum_{n} \sqrt{\tilde{\Lambda}_{n}}
                        u_{n}(\theta)w_{n}^{*}(\theta'),
                 \label{eq:joint_sdecomp}
            \end{split}
            \end{equation}
where $\Lambda_{n}$ and $\tilde{\Lambda}_{n}=\Lambda_n +1$ are the eigenvalues and $u_n(\theta)$ and $w_n(\theta')$ the eigenfunctions. One can parameterize the eigenvalues as $ \Lambda_n = \sinh^2{(r_n)} $ and $\tilde{\Lambda}_{n}=\cosh^2{(r_n)}$, where the parameter $r_n$ can be interpreted as a gain in mode $n$. 

Using Eq. (\ref{eq:sol_ops_first_crystal}), we can calculate the angular intensity distribution:
\begin{equation}
\langle  \hat{N}(\theta) \rangle =\langle \left[\hat{a}^{\mathrm{out}}(\theta)\right]^\dagger \hat{a}^{\mathrm{out}}(\theta) \rangle =  \int d\theta'\,|\beta (\theta,\theta')|^2,
 \label{intensity}
 \end{equation}
which, due to the orthogonality of the Schmidt modes, can be represented as the sum of the intensities of individual Schmidt modes:
\begin{equation}
\langle \hat{N}(\theta) \rangle =  \sum_n \Lambda_{n} |u_n(\theta)|^2.
 \label{intensity_Schmidt}
 \end{equation}
The gain $G_2$ of the amplifier, which we measure (see the SI) from the pump power dependence of the mean photon number at zero angle, $\langle \hat{N}^{\mathrm{AV}}(\theta=0)\rangle= \sinh^2(G_2)$, is then related to the gain values of the Schmidt modes:
  \begin{equation}
   \label{equation}
  \langle \hat{N}^{\mathrm{AV}}(\theta=0) \rangle = \sum_n \Lambda_n |u_n(\theta=0)|^2 . 
  \end{equation} 
  
Using the Schmidt decomposition, we introduce the input and output broadband Schmidt operators:
           \begin{equation}
         \hat{A}_n^{\mathrm{in}}= \int d\theta  \  w^*_n(\theta) \hat{a}^{\mathrm{in}} (\theta),\, \ \ \
                  \hat{A}_n^{\mathrm{out}}= \int d\theta \   u^*_n(\theta) \hat{a}^{\mathrm{out}} (\theta).
        \end{equation}

             We see that the input and output Schmidt operators are related to different eigenfunctions, $w_n(\theta)$ and $ u_n(\theta)$, respectively. Therefore, we will further call the set of functions  $w_n(\theta)$ {\it the input Schmidt modes}, 
             while the functions  $u_n(\theta)$, {\it the output Schmidt modes}. These eigenfunctions have the same absolute values but different phases. The phase difference is a consequence of the time-ordering effect, which is important in the high-gain regime~\cite{Christ2013May}, but can be neglected in the low-gain regime, where $w_n(\theta) = u_n(\theta)$.

         Substituting the Schmidt decomposition  (\ref{eq:joint_sdecomp}) into the Bogoliubov transformation for the plane-wave operators  (\ref{eq:sol_ops_first_crystal}), we obtain the input/output relations for the Schmidt operators,
\begin{equation}
     \label{eq:sol_ops_Schmidt}
\hat{A}_n^{\mathrm{out}} =   \sqrt{\Lambda_{n}+1} \hat{A}_n^{\mathrm{in}}
                    +    \sqrt{\Lambda_{n}} \left[\hat{A}_n^{\mathrm{in}}\right]^{\dagger},       
       \end{equation}    
          while the inverse transformation reads
       \begin{equation}
\hat{A}_n^{\mathrm{in}} =   \sqrt{\Lambda_{n}+1}  \hat{A}_n^{\mathrm{out}}
                    -   \sqrt{\Lambda_{n}} \left[\hat{A}_n^{\mathrm{out}}\right]^{\dagger}.
              \label{eq:inv_Bog}
        \end{equation}

The output Schmidt operators of the first crystal (squeezer) $\hat{A}_l$ can be decomposed over the input Schmidt operators of the second crystal (amplifier) $ \hat{A}_n^{\mathrm{amp,in}}$:
\begin{equation}
\label{decomposition_1_2}
         \hat{A}_l = \sum_n g_{ln} \ \hat{A}_n^{\mathrm{amp,in}},
         \end{equation}
where $g_{ln} = \int d\theta  \ \left[ v_l (\theta)\right]^*  \  \ w_n (\theta) $ are the overlap coefficients between the output modes $v_l (\theta)$ of the squeezer  and the input modes $w_n (\theta)$ of the amplifier. Note that according to the Schmidt decomposition, both modes $v_n(\theta)$ and  $w_n(\theta)$ form orthogonal sets, therefore $\sum_n |g_{ln}|^2 = \sum_l |g_{ln}|^2 =1$.

Similarly, the output Schmidt operators of the amplifier $\hat{A}_n^{ \mathrm{amp,out}}$ can be decomposed over the output Schmidt operators of the entire  interferometer $\hat{A}_k^{\mathrm{int,out}}$: 
\begin{equation}
\label{decomposition_2_SU}
         \hat{A}_n^{\mathrm{amp,out}} = \sum_k h_{nk} \hat{A}_k^{\mathrm{int,out}},
         \end{equation}
where $h_{nk} = \int d\theta  \ \left[u^{\mathrm{amp}}_n (\theta)\right]^*  u_k (\theta) $ are the corresponding overlap coefficients with the normalization $\sum_n |h_{nk}|^2 = \sum_k |h_{nk}|^2 =1$. Note that the output modes of the amplifier $u^{\mathrm{amp}}_n(\theta)$ coincide with the amplified vacuum modes mentioned in the main text. 

Substituting  Eq. (\ref{eq:inv_Bog}) and  Eq. (\ref{decomposition_2_SU}) in Eq. (\ref{decomposition_1_2}), we find a connection between the output Schmidt operators of the squeezer and the output Schmidt operators of the entire interferometer: 
 \begin{equation}
 \label{eq:connection_relation}
  \hat{A}_l = \sum_{n,k} g_{ln} \left[\sqrt{\Lambda_n +1 } \ h_{nk}   \hat{A}_k^{\mathrm{int,out}} \mp \sqrt{\Lambda_n } \ h^*_{nk}  \left[\hat{A}_k ^{\mathrm{int,out}}\right]^{\dagger} \right],
 \end{equation}
where the $"-"$ sign corresponds to the positive gain of the amplifier (resulting in the bright fringe), while $"+"$ corresponds to the negative gain of the amplifier (resulting in the dark fringe).

\section{Appendix B: The role of spherical mirror alignment}
\label{App2}

From Eq.~(\ref{eq:connection_relation}), we express the quadrature operators $\hat{Q}_l$ and $\hat{P}_l$ of the squeezer mode $l$ in terms of the quadrature operators of the interferometer output modes $n$.
Denoting the output  quadrature operators at the bright and dark fringes as $\hat{Q}^{\mathrm{B}}_n, \hat{P}^{\mathrm{B}}_n$ and $\hat{Q}^{\mathrm{D}}_n, \hat{P}^{\mathrm{D}}_n$, respectively,  we obtain 
\begin{eqnarray}
 \label{eq:B/D}
 \hat{Q}_l &=& \sum_{n} \left[ \mathrm{Re}[{M}_{ln}^{-/+}] \hat{Q}^{\mathrm{B/D}}_n-\mathrm{Im}[{M}_{ln}^{-/+}] \hat{P}^{\mathrm{B/D}}_n\right],\nonumber\\
 \hat{P}_l &=& \sum_{n} \left[ \mathrm{Im}[{M}_{ln}^{+/-}] \hat{Q}^{\mathrm{B/D}}_n+\mathrm{Re}[{M}_{ln}^{+/-}] \hat{P}^{\mathrm{B/D}}_n\right],
 \end{eqnarray}
 where we introduced the matrices ${M}_{ln}^{-/+} = g_{ln}\sqrt{\Lambda_n+1} \mp g^*_{ln}\sqrt{\Lambda_n} $.
  
We align the spherical mirror so that the intensity distribution at the interferometer output (Fig.~\ref{SqEffandAI}b) oscillates as a whole, with the intensity at each point being minimal for the dark fringe and maximal for the bright fringe. This means that for each mode, the $\hat{Q}_l$ quadrature (already anti-squeezed, i.e., amplified) is further amplified in the bright fringe and de-amplified in the dark fringe, 
while the $\hat{P}_l$ (squeezed) quadrature behaves oppositely: it is amplified in the dark fringe and de-amplified in the bright fringe (see Fig.~S1 of the SI for more details). This condition means that for each mode $l$, in the bright fringe the amplified quadratures ($Q_n^{\mathrm{B}}$) get input only from quadratures $Q_l$, while in the dark fringe, the amplified quadratures ($P_n^{\mathrm{D}}$) get input only from quadratures $P_l$. It is achieved when $\mathrm{Im}[M^{+}_{ln}]=\mathrm{Im}[M^{-}_{ln}]=0$.

As a result, the alignment procedure balances the phases of the modes, ensuring that the $g_{ln}$ matrix is real. Consequently, Eqs.~(\ref{eq:B/D}) lead to simple relations between the quadrature operators of mode $l$ of the squeezed vacuum   and the quadrature operators of all modes at the interferometer output:
\begin{equation}
\label{eq:simp}
\begin{split}
 \hat{Q}_l = \sum_{n} g_{ln}(\sqrt{\Lambda_n+1}-/+ \sqrt{\Lambda_n}) \hat{Q}^{\mathrm{B/D}}_n,
 \\
 \hat{P}_l = \sum_{n} g_{ln}(\sqrt{\Lambda_n+1}+/-\sqrt{\Lambda_n}) \hat{P}^{\mathrm{B/D}}_n.
 \end{split}
 \end{equation}

\begin{backmatter}
\bmsection{Funding} This work was funded within the QuantERA
II Programme (project SPARQL) that has received funding from the
European Union’s Horizon 2020 research and innovation programme
 under Grant Agreement No 101017733, with the funding organization
 Deutsche Forschungsgemeinschaft. The project/research is also part of the Munich Quantum Valley, which is supported by the Bavarian state government with funds from the Hightech Agenda Bavaria. I. B., M. K., N. L., and M. C. are part of the Max Planck School of Photonics, supported by BMBF, Max Planck Society,
 and Fraunhofer Society.
 We acknowledge financial support of the Deutsche Forschungsgemeinschaft (DFG) via Project SH 1228/3-1 and via the TRR 142/3 (Project No. 231447078, Subproject No. C10). We also thank the PC2 (Paderborn Center for Parallel Computing) for providing computation time.


\smallskip

\bmsection{Disclosures} The authors declare no conflicts of interest.

\bmsection{Supplementary}
https://www.overleaf.com/project/624eabe3db66bdf5c957d150

\bmsection{Data availability} Data underlying the results presented in this paper are not publicly available at this time but may be obtained from the authors upon reasonable request.

\end{backmatter}
\bibliography{MainText}

\end{document}